# Linear Algebra and Number of Spanning Trees

E. M. Badr

Scientific Computing Department, Faculty of Computers and Informatics
Benha University, Egypt.
E-mail: badrgraph@gmail.com

B. Mohamed
Mathematics Department, Faculty of Science
Menoufia University, Egypt.

**Abstract:**

A network-theoretic approach for determining the complexity of a graph is proposed. This approach is based on the relationship between the linear algebra (theory of determinants) and the graph theory. In this paper we contribute a new algebraic method to derive simple formulas of the complexity of some new networks using linear algebra. We apply this method to derive the explicit formulas for the friendship network $F_3^k$ and the subdivision of friendship graph $S(F_3^k)$. We also calculate their spanning trees entropy and compare it between them. Finally, we introduce an open problem "Any improvement for calculating of the determinant in linear algebra, we can investigate this improvement as a new method to determine the number of spanning tree for a given graph.

## 1- Introduction:

We deal with simple and finite graphs $G = (V, E)$, where V is the vertex set and $E$ is the edge set. A graph $T$ is called a tree if it has not circuits so there is exactly one path from each vertex in the tree to each other vertex in the tree. Spanning tree of a graph $G$ is a tree containing all vertices of $G$. The number of spanning trees in $G$, also called, the complexity of the graph G, denoted by $\tau(G)$.

There are many methods which determine the number of spanning trees for a given graph $G$. Some of them are classified under combinatorial approach [1, 2, 3, 4, 5] and the other are classified under algebraic approach [6, 7, 8, 9].
Chebyshev polynomials method was used to determine the number of spanning trees for a given graph $G$ as an algebraic method [10, 11, 12, 13, 14].

In this paper we contribute a new algebraic method which is based on Dodgson and Chio's method. It has an advantage that calculate determinants of $n \times n$ ($n \geq 3$) matrix, by reducing determinants to 2$^{nd}$ Order.

**2- Dodgson and Chio's condensation method**

Chio's condensation is a method for evaluating an $n \times n$ determinant in terms of $(n-1) \times (n-1)$ determinants ; see [15]:

$$A = \begin{vmatrix} a_{11} & a_{12} & \cdots & a_{1n} \\ a_{21} & a_{22} & \cdots & a_{2n} \\ \vdots & \vdots & \ddots & \vdots \\ a_{n1} & a_{n2} & \cdots & a_{nn} \end{vmatrix} = \frac{1}{a_{11}^{n-2}} \begin{vmatrix} \begin{vmatrix} a_{11} & a_{12} \\ a_{21} & a_{22} \end{vmatrix} & \begin{vmatrix} a_{11} & a_{13} \\ a_{21} & a_{23} \end{vmatrix} & \cdots & \begin{vmatrix} a_{11} & a_{1n} \\ a_{21} & a_{2n} \end{vmatrix} \\ \begin{vmatrix} a_{11} & a_{12} \\ a_{31} & a_{32} \end{vmatrix} & \begin{vmatrix} a_{11} & a_{13} \\ a_{31} & a_{33} \end{vmatrix} & \cdots & \begin{vmatrix} a_{11} & a_{1n} \\ a_{31} & a_{3n} \end{vmatrix} \\ \vdots & \vdots & \ddots & \vdots \\ \begin{vmatrix} a_{11} & a_{12} \\ a_{n1} & a_{n2} \end{vmatrix} & \begin{vmatrix} a_{11} & a_{13} \\ a_{n1} & a_{n3} \end{vmatrix} & \cdots & \begin{vmatrix} a_{11} & a_{1n} \\ a_{n1} & a_{nn} \end{vmatrix} \end{vmatrix}$$

Dodgson's condensation method computes determinants of size n×n by expressing them in terms of those of size $(n-1) \times (n-1)$, and then expresses the latter in terms of determinants of size $(n-2) \times (n-2)$, and so on (see [16]).

Armend [17] proposed another method is based on Dodgson and Chio's method, but the difference between them is that this new method is resolved by calculating 4 unique determinants of $(n-1) \times (n-1)$ Order, (which can be derived from determinants of $n \times n$ order, if we remove first row and first column or first row and last column or last row and first column or last row and last column, elements that belongs to only one of unique determinants we should call them unique elements), and one determinant of $(n-2) \times (n-2)$ order which is formed from $n \times n$ order determinant with elements $a_{i,j}$ with $i, j \neq 1$, n, on condition that the determinant of $(n-2) \times (n-2) \neq 0$.

**Theorem 3.1 [17]:**

Every determinant of $n \times n$ (n > 2) order can be reduced into 2×2 order determinant, by calculating 4 determinants of $(n-1) \times (n-1)$ order, and one determinant of $(n-2) \times (n-2)$ order, on condition that $(n-2) \times (n-2)$ order determinants to be different from zero.

Ongoing is presented a scheme of calculating the determinants of $n \times n$ order according to this formula:

$$A = \begin{vmatrix} a_{11} & a_{12} & \cdots & a_{1n} \\ a_{21} & a_{22} & \cdots & a_{2n} \\ \vdots & \vdots & \ddots & \vdots \\ a_{n1} & a_{n2} & \cdots & a_{nn} \end{vmatrix} = \frac{1}{|B|} \cdot \begin{vmatrix} |C| & |D| \\ |E| & |F| \end{vmatrix}, |B| \neq 0$$

The $|B|$ *is* $(n-2) \times (n-2)$ order determinant which is the interior determinant of determinant $|A|$ while $|C|$, $|D|$, $|E|$ and $|F|$ are unique determinants of $(n-1) \times (n-1)$ order, which can be formed from $n \times n$ order determinant.

**Proof**: We can see [17].

**Theorem 3.2:** The number of spanning trees of the friendship graph $F_3^k$

$\tau(F_3^k) = 3^k, k \geq 1,$ where $k$ is the number of blocks.

**Proof :**

Let p = 2k+1 and q = 3k are the number of vertices and edges of $F_3^k$ respectively.

Let the $L = D - A$ be the Laplecian matrix for $F_3^k$ such that :

$$L = \begin{vmatrix} 2k & -1 & -1 & \cdots & \cdots & \cdots & \cdots & -1 \\ -1 & 2 & -1 & 0 & \cdots & \cdots & \cdots & 0 \\ -1 & -1 & 2 & 0 & \cdots & \cdots & \cdots & 0 \\ \vdots & 0 & 0 & \ddots & \ddots & 0 & \cdots & 0 \\ \vdots & \vdots & \vdots & \ddots & \ddots & \ddots & \cdots & \vdots \\ \vdots & \vdots & \vdots & \ddots & \ddots & \ddots & -1 & \vdots \\ \vdots & \vdots & \vdots & \ddots & \ddots & -1 & 2 & 0 \\ \vdots & 0 & \cdots & \cdots & \cdots & 0 & 0 & 2 \end{vmatrix}_{n \times n}$$

According to Kirchhoff' Theorem, the number of spanning trees for $F_3^k$ is the co-factor matrix L that is:

$$\tau(F_3^k) = \begin{vmatrix} 2 & -1 & 0 & \cdots & \cdots & \cdots & 0 \\ -1 & 2 & 0 & \cdots & \cdots & \cdots & 0 \\ 0 & 0 & \ddots & \ddots & 0 & \cdots & 0 \\ \vdots & \vdots & \ddots & \ddots & \ddots & 0 & 0 \\ \vdots & \vdots & \ddots & \ddots & \ddots & -1 & \vdots \\ \vdots & \vdots & \ddots & \ddots & \ddots & 2 & \vdots \\ 0 & \cdots & \cdots & \cdots & 0 & 0 & 2 \end{vmatrix}_{(n-1) \times (n-1)}$$

According to Dodgson and Chio's method, we have

$$\tau(F_3^k) = \begin{vmatrix} 2 & -1 & 0 & \cdots & \cdots & \cdots & 0 \\ -1 & 2 & 0 & \cdots & \cdots & \cdots & 0 \\ 0 & 0 & \ddots & \ddots & 0 & \cdots & 0 \\ \vdots & \vdots & \ddots & \ddots & \ddots & 0 & 0 \\ \vdots & \vdots & \ddots & \ddots & \ddots & -1 & \vdots \\ \vdots & \vdots & \ddots & \ddots & \ddots & 2 & \vdots \\ 0 & \cdots & \cdots & \cdots & 0 & 0 & 2 \end{vmatrix} = \frac{1}{|B|} \begin{vmatrix} |C| & |D| \\ |D^T| & |E| \end{vmatrix} = 3^k, |B| \neq 0 \quad \text{such that :}$$

$$B = \begin{bmatrix} 2 & -1 & 0 & \cdots & \cdots & 0 \\ -1 & \ddots & 0 & \vdots & \vdots & \vdots \\ 0 & \cdots & \ddots & -1 & \vdots & \vdots \\ \vdots & \vdots & -1 & \ddots & \vdots & 0 \\ \vdots & \vdots & \cdots & \cdots & \ddots & -1 \\ 0 & \cdots & \cdots & 0 & -1 & 2 \end{bmatrix}$$

$$C = \begin{bmatrix} 2S & -1 & \cdots & \cdots & \cdots & \cdots & -1 \\ -1 & 2 & -1 & 0 & \cdots & \cdots & 0 \\ \vdots & -1 & \ddots & \vdots & \cdots & \cdots & \vdots \\ \vdots & 0 & \cdots & \ddots & -1 & \vdots & \vdots \\ \vdots & \vdots & \vdots & -1 & \ddots & \vdots & 0 \\ \vdots & \vdots & \vdots & \cdots & \cdots & \ddots & -1 \\ -1 & 0 & \cdots & \cdots & 0 & -1 & 2 \end{bmatrix}$$

$$D = \begin{bmatrix} -1 & \cdots & \cdots & \cdots & \cdots & \cdots & -1 \\ 2 & -1 & 0 & \cdots & \cdots & \cdots & 0 \\ -1 & \ddots & \vdots & \cdots & \cdots & \cdots & \vdots \\ 0 & 0 & \ddots & -1 & \vdots & \vdots & \vdots \\ \vdots & \vdots & -1 & \ddots & \vdots & \vdots & \vdots \\ \vdots & \vdots & \cdots & \cdots & \ddots & -1 & \vdots \\ 0 & \cdots & \cdots & 0 & -1 & 2 & 0 \end{bmatrix}$$

$$E = D^T = \begin{bmatrix} -1 & 2 & -1 & 0 & \cdots & \cdots & 0 \\ \vdots & -1 & 2 & \vdots & \cdots & \cdots & \vdots \\ \vdots & 0 & 0 & \ddots & -1 & \vdots & \vdots \\ \vdots & \vdots & \vdots & -1 & \ddots & \vdots & 0 \\ \vdots & \vdots & \vdots & \vdots & \cdots & \cdots & -1 \\ \vdots & \vdots & \vdots & \vdots & \cdots & -1 & 2 \\ -1 & 0 & \cdots & \cdots & \cdots & \cdots & 0 \end{bmatrix}$$

$$F = \begin{bmatrix} 2 & -1 & 0 & \cdots & \cdots & \cdots & 0 \\ -1 & \ddots & \vdots & \cdots & \cdots & \cdots & \vdots \\ 0 & 0 & \ddots & -1 & \vdots & \vdots & \vdots \\ \vdots & \vdots & \ddots & \ddots & \vdots & \vdots & \vdots \\ \vdots & \vdots & 0 & \ddots & \ddots & -1 & \vdots \\ \vdots & \vdots & \vdots & \cdots & -1 & \ddots & 0 \\ 0 & \cdots & \cdots & \cdots & \cdots & 0 & 2 \end{bmatrix}$$

where B, C, D, E, F are $(2k-2)\times(2k-2), (2k-1)\times(2k-1), (2k-1)\times(2k-1)$, $(2k-1)\times(2k-1), (2k-1)\times(2k-1)$ respectively.

By using properties of determinants we get:

$$\therefore |B| = \begin{vmatrix} 2 & 0 & \cdots & \cdots & \cdots & 0 \\ 0 & \frac{3}{2} & \ddots & \ddots & \ddots & \vdots \\ \vdots & \ddots & \ddots & \ddots & \ddots & \vdots \\ \vdots & \ddots & \ddots & \ddots & \ddots & \vdots \\ \vdots & \ddots & \ddots & \ddots & 2 & 0 \\ 0 & \cdots & \cdots & \cdots & 0 & \frac{3}{2} \end{vmatrix} = 3^{S-1}$$

$$, |C| = \begin{vmatrix} 2S & -1 & \cdots & \cdots & \cdots & \cdots & -1 \\ -1 & 2 & -1 & 0 & \cdots & \cdots & 0 \\ \vdots & -1 & \ddots & \vdots & \cdots & \cdots & \vdots \\ \vdots & 0 & \cdots & \ddots & -1 & \vdots & \vdots \\ \vdots & \vdots & \vdots & -1 & \ddots & \vdots & 0 \\ \vdots & \vdots & \vdots & \cdots & \cdots & \ddots & -1 \\ -1 & 0 & \cdots & \cdots & 0 & -1 & 2 \end{vmatrix} = 2*3^{S-1}$$

$$,|E|=|D^T|=|D|=\begin{vmatrix} 2 & 0 & \cdots & \cdots & \cdots & 0 \\ 0 & \frac{-3}{2} & \ddots & \ddots & \ddots & \vdots \\ \vdots & \ddots & \ddots & \ddots & \ddots & \vdots \\ \vdots & \ddots & \ddots & \ddots & \ddots & \vdots \\ \vdots & \ddots & \ddots & \ddots & 2 & \ddots & \vdots \\ \vdots & \ddots & \ddots & \ddots & \ddots & \frac{-3}{2} & 0 \\ 0 & \cdots & \cdots & \cdots & 0 & -1 \end{vmatrix}=-3^{S-1} \quad |E|=\begin{vmatrix} 2 & 0 & \cdots & \cdots & \cdots & 0 \\ 0 & \frac{3}{2} & \ddots & \ddots & \ddots & \vdots \\ \vdots & \ddots & \ddots & \ddots & \ddots & \vdots \\ \vdots & \ddots & \ddots & \ddots & 2 & \ddots & \vdots \\ \vdots & \ddots & \ddots & \ddots & \ddots & \frac{3}{2} & 0 \\ 0 & \cdots & \cdots & \cdots & 0 & 2 \end{vmatrix}=2*3^{S-1}$$

Therefore we get

$$\tau(F_S)=\frac{1}{|B|}\begin{vmatrix} |C| & |D| \\ |D^T| & |E| \end{vmatrix}=\frac{1}{3^{k-1}}\begin{vmatrix} 2*3^{k-1} & -3^{k-1} \\ -3^{k-1} & 2*3^{k-1} \end{vmatrix}=3^k, k\geq 2. \qquad \blacksquare$$

**Theorem 3.3:** The number of spanning trees of the subdivision of friendship graph $S(F_3^k)$ is $\tau(S(F_3^k))=6^k, k\geq 1$, where k is the number of blocks.

**Proof:**
Let p = 5k + 1 and q = 6k are the number of vertices and edges of $S(F_3^k)$ respectively.
Let the $L = D - A$ be the Laplecian matrix for $S(F_3^k)$ such that:

$$L=\begin{bmatrix}
2k & -1 & -1 & 0 & 0 & 0 & -1 & -1 & 0 & 0 & \cdots & \cdots & \cdots & \cdots & \cdots & 0 & -1 & -1 & 0 & 0 \\
-1 & 2 & 0 & -1 & 0 & \cdots & \cdots & \cdots & \cdots & \cdots & \cdots & \cdots & \cdots & \cdots & \cdots & \cdots & \cdots & \cdots & 0 & 0 \\
-1 & 0 & 2 & 0 & 0 & -1 & 0 & \cdots & \cdots & \cdots & \cdots & \cdots & \cdots & \cdots & \cdots & \cdots & \cdots & \cdots & \cdots & \vdots \\
0 & -1 & 0 & 2 & -1 & 0 & \cdots & \cdots & \cdots & \cdots & \cdots & \cdots & \cdots & \cdots & \cdots & \cdots & \cdots & \cdots & \cdots & \vdots \\
0 & 0 & 0 & -1 & 2 & -1 & 0 & \cdots & \cdots & \cdots & \cdots & \cdots & \cdots & \cdots & \cdots & \cdots & \cdots & \cdots & \cdots & \vdots \\
0 & \vdots & -1 & 0 & -1 & 2 & 0 & \cdots & \cdots & \cdots & \cdots & \cdots & \cdots & \cdots & \cdots & \cdots & \cdots & \cdots & \cdots & \vdots \\
-1 & \vdots & 0 & \vdots & 0 & 0 & 2 & \ddots & -1 & 0 & \cdots & \cdots & \cdots & \cdots & \cdots & \cdots & \cdots & \cdots & \cdots & \vdots \\
-1 & \vdots & \vdots & \vdots & \vdots & \vdots & \ddots & 2 & \ddots & 0 & -1 & 0 & \cdots & \cdots & \cdots & \cdots & \cdots & \cdots & \cdots & \vdots \\
0 & \vdots & \vdots & \vdots & \vdots & \vdots & -1 & \ddots & 2 & -1 & 0 & \cdots & \cdots & \cdots & \cdots & \cdots & \cdots & \cdots & \cdots & \vdots \\
0 & \vdots & \vdots & \vdots & \vdots & \vdots & 0 & 0 & -1 & 2 & -1 & 0 & \cdots & \cdots & \cdots & \cdots & \cdots & \cdots & \cdots & \vdots \\
\vdots & \vdots & \vdots & \vdots & \vdots & \vdots & \vdots & -1 & \ddots & -1 & \ddots & \vdots & \cdots & \cdots & \cdots & \cdots & \cdots & \cdots & \cdots & \vdots \\
\vdots & \vdots & \vdots & \vdots & \vdots & \vdots & \vdots & 0 & \vdots & \ddots & 0 & \ddots & \ddots & -1 & 0 & \cdots & \cdots & \cdots & \cdots & \vdots \\
\vdots & \vdots & \vdots & \vdots & \vdots & \vdots & \vdots & \vdots & \vdots & 0 & \ddots & \ddots & \ddots & 0 & -1 & 0 & \cdots & \cdots & \cdots & \vdots \\
\vdots & \vdots & \vdots & \vdots & \vdots & \vdots & \vdots & \vdots & \vdots & \vdots & \ddots & -1 & \ddots & \ddots & -1 & 0 & \cdots & \cdots & \cdots & \vdots \\
\vdots & \vdots & \vdots & \vdots & \vdots & \vdots & \vdots & \vdots & \vdots & \vdots & \ddots & \ddots & 0 & -1 & \ddots & -1 & 0 & \cdots & \cdots & \vdots \\
0 & \vdots & \vdots & \vdots & \vdots & \vdots & \vdots & \vdots & \vdots & \vdots & \cdots & \cdots & -1 & 0 & -1 & 2 & 0 & \cdots & \cdots & \cdots \\
-1 & \vdots & \vdots & \vdots & \vdots & \vdots & \vdots & \vdots & \vdots & \vdots & \cdots & \cdots & \cdots & \cdots & \cdots & 0 & 2 & \ddots & -1 & 0 \\
-1 & \vdots & \vdots & \vdots & \vdots & \vdots & \vdots & \vdots & \vdots & \vdots & \cdots & \cdots & \cdots & \cdots & \cdots & \vdots & \ddots & 2 & \ddots & 0 \\
0 & 0 & \vdots & \vdots & \vdots & \vdots & \vdots & \vdots & \vdots & \vdots & \cdots & \cdots & \cdots & \cdots & \cdots & \vdots & -1 & \ddots & 2 & -1 \\
0 & 0 & \vdots & \vdots & \vdots & \vdots & \vdots & \vdots & \vdots & \vdots & \cdots & \cdots & \cdots & \cdots & \cdots & \vdots & \cdots & 0 & -1 & 2
\end{bmatrix}_{n\times n}$$

According to Kirchhoff' Theorem, the number of spanning trees for $F_3^k$ is the cofactor matrix L that is:

$$\tau(F_3^k) = \begin{vmatrix} 2 & 0 & -1 & 0 & \cdots & \cdots & \cdots & \cdots & \cdots & \cdots & \cdots & \cdots & \cdots & \cdots & \cdots & \cdots & 0 & 0 \\ 0 & 2 & 0 & 0 & -1 & 0 & \cdots & \cdots & \cdots & \cdots & \cdots & \cdots & \cdots & \cdots & \cdots & \cdots & \cdots & \vdots \\ -1 & 0 & 2 & -1 & 0 & \cdots & \cdots & \cdots & \cdots & \cdots & \cdots & \cdots & \cdots & \cdots & \cdots & \cdots & \cdots & \vdots \\ 0 & 0 & -1 & 2 & -1 & 0 & \cdots & \cdots & \cdots & \cdots & \cdots & \cdots & \cdots & \cdots & \cdots & \cdots & \cdots & \vdots \\ \vdots & -1 & 0 & -1 & 2 & 0 & \cdots & \cdots & \cdots & \cdots & \cdots & \cdots & \cdots & \cdots & \cdots & \cdots & \cdots & \vdots \\ \vdots & 0 & \vdots & 0 & 0 & 2 & \ddots & -1 & 0 & \cdots & \cdots & \cdots & \cdots & \cdots & \cdots & \cdots & \cdots & \vdots \\ \vdots & \vdots & \vdots & \vdots & \vdots & \ddots & 2 & \ddots & 0 & -1 & 0 & \cdots & \cdots & \cdots & \cdots & \cdots & \cdots & \vdots \\ \vdots & \vdots & \vdots & \vdots & \vdots & -1 & \ddots & 2 & -1 & 0 & \cdots & \cdots & \cdots & \cdots & \cdots & \cdots & \cdots & \vdots \\ \vdots & \vdots & \vdots & \vdots & \vdots & 0 & 0 & -1 & 2 & -1 & 0 & \cdots & \cdots & \cdots & \cdots & \cdots & \cdots & \vdots \\ \vdots & \vdots & \vdots & \vdots & \vdots & \vdots & -1 & \ddots & -1 & \ddots & \vdots & \cdots & \cdots & \cdots & \cdots & \cdots & \cdots & \vdots \\ \vdots & \vdots & \vdots & \vdots & \vdots & \vdots & 0 & \vdots & \ddots & 0 & \ddots & \ddots & -1 & 0 & \cdots & \cdots & \cdots & \vdots \\ \vdots & \vdots & \vdots & \vdots & \vdots & \vdots & \vdots & \vdots & 0 & \ddots & \ddots & \ddots & 0 & -1 & 0 & \cdots & \cdots & \vdots \\ \vdots & \vdots & \vdots & \vdots & \vdots & \vdots & \vdots & \vdots & \ddots & -1 & \ddots & \ddots & -1 & 0 & \cdots & \cdots & \cdots & \vdots \\ \vdots & \vdots & \vdots & \vdots & \vdots & \vdots & \vdots & \vdots & \ddots & \ddots & 0 & -1 & \ddots & -1 & 0 & \cdots & \cdots & \vdots \\ \vdots & \vdots & \vdots & \vdots & \vdots & \vdots & \vdots & \vdots & \cdots & \cdots & -1 & 0 & -1 & 2 & 0 & \cdots & \cdots & \cdots \\ \vdots & \vdots & \vdots & \vdots & \vdots & \vdots & \vdots & \vdots & \cdots & \cdots & \cdots & \cdots & 0 & 2 & \ddots & -1 & 0 \\ \vdots & \vdots & \vdots & \vdots & \vdots & \vdots & \vdots & \vdots & \cdots & \cdots & \cdots & \cdots & \cdots & \vdots & \ddots & 2 & \ddots & 0 \\ 0 & \vdots & \vdots & \vdots & \vdots & \vdots & \vdots & \vdots & \cdots & \cdots & \cdots & \cdots & \cdots & \vdots & -1 & \ddots & 2 & -1 \\ 0 & \vdots & \vdots & \vdots & \vdots & \vdots & \vdots & \vdots & \cdots & \cdots & \cdots & \cdots & \cdots & \vdots & \cdots & 0 & -1 & 2 \end{vmatrix}_{(n-1)\times(n-1)}$$

According to Dodgson and Chio's method, we have

$$\tau(S(F_k)) = \frac{1}{|B|}\begin{Vmatrix} |C| & |D| \\ |D^T| & |E| \end{Vmatrix} = 6^k, |B| \neq 0$$

Such that:

$$B = \begin{bmatrix} 2 & 0 & -1 & 0 & \cdots & \cdots & \cdots & 0 \\ 0 & \ddots & 0 & 0 & -1 & 0 & \cdots & \vdots \\ -1 & 0 & \ddots & -1 & 0 & \cdots & \cdots & \vdots \\ 0 & 0 & -1 & \ddots & -1 & \vdots & \vdots & \vdots \\ \vdots & -1 & 0 & -1 & \ddots & \vdots & \vdots & 0 \\ \vdots & 0 & \vdots & 0 & 0 & \ddots & 0 & -1 \\ \vdots & \vdots & \vdots & \vdots & \vdots & 0 & \ddots & 0 \\ 0 & \cdots & \cdots & \cdots & 0 & -1 & 0 & 2 \end{bmatrix}$$

$$C = \begin{bmatrix} 2S & -1 & -1 & 0 & \cdots & -1 & -1 & 0 & \cdots \\ -1 & 2 & 0 & -1 & 0 & \cdots & \cdots & \cdots & 0 \\ -1 & 0 & \ddots & 0 & 0 & -1 & 0 & \cdots & \vdots \\ 0 & -1 & 0 & \ddots & -1 & 0 & \cdots & \cdots & \vdots \\ \vdots & 0 & 0 & -1 & \ddots & -1 & 0 & \cdots & 0 \\ 0 & 0 & -1 & 0 & -1 & \ddots & 0 & \cdots & 0 \\ -1 & \vdots & 0 & \cdots & \cdots & 0 & \ddots & 0 & -1 \\ -1 & \vdots & \vdots & \cdots & \cdots & \cdots & 0 & \ddots & 0 \\ 0 & \cdots & \cdots & \cdots & \cdots & \cdots & 0 & -1 & 0 & 2 \end{bmatrix}$$

$$D = \begin{bmatrix} -1 & -1 & 0 & \cdots & \cdots & -1 & -1 & 0 & 0 \\ 2 & 0 & -1 & 0 & 0 & 0 & 0 & \vdots & \vdots \\ 0 & \ddots & 0 & 0 & -1 & 0 & \vdots & \vdots & \vdots \\ -1 & 0 & \ddots & -1 & 0 & 0 & \vdots & \vdots & \vdots \\ 0 & 0 & -1 & \ddots & -1 & 0 & \vdots & \vdots & \vdots \\ \vdots & -1 & 0 & -1 & \ddots & 0 & \vdots & 0 & \vdots \\ \vdots & 0 & \vdots & 0 & 0 & \ddots & 0 & -1 & \vdots \\ \vdots & \vdots & \vdots & \vdots & \vdots & 0 & \ddots & 0 & 0 \\ 0 & \cdots & \cdots & \cdots & 0 & -1 & 0 & 2 & -1 \end{bmatrix}$$

$$E = D^T = \begin{bmatrix} -1 & 2 & 0 & -1 & 0 & \cdots & \cdots & \cdots & 0 \\ -1 & 0 & 2 & 0 & 0 & -1 & 0 & \cdots & \vdots \\ 0 & -1 & 0 & 2 & -1 & 0 & \cdots & \cdots & \vdots \\ 0 & 0 & 0 & -1 & \ddots & -1 & 0 & \cdots & \vdots \\ 0 & \vdots & -1 & 0 & -1 & \ddots & 0 & \cdots & 0 \\ -1 & \vdots & 0 & \vdots & \cdots & \cdots & \ddots & 0 & -1 \\ -1 & \vdots & \vdots & \vdots & \cdots & \cdots & \cdots & \ddots & 0 \\ 0 & \vdots & \vdots & \vdots & \cdots & 0 & -1 & 0 & 2 \\ \vdots & \cdots & \cdots & \cdots & \cdots & \cdots & \cdots & 0 & -1 \end{bmatrix}$$

$$F = \begin{bmatrix} 2 & 0 & -1 & 0 & \cdots & \cdots & \cdots & \cdots & 0 \\ 0 & 2 & 0 & 0 & -1 & 0 & \cdots & \cdots & \vdots \\ -1 & 0 & \ddots & -1 & 0 & \cdots & \cdots & \cdots & \vdots \\ 0 & 0 & -1 & \ddots & -1 & 0 & \cdots & \cdots & \vdots \\ \vdots & -1 & 0 & -1 & \ddots & 0 & \cdots & \cdots & \vdots \\ \vdots & \cdots & \cdots & \cdots & \cdots & \ddots & 0 & -1 & \vdots \\ \vdots & \cdots & \cdots & \cdots & \cdots & \cdots & \ddots & 0 & 0 \\ \vdots & \cdots & \cdots & \cdots & 0 & -1 & 0 & \ddots & -1 \\ 0 & \cdots & \cdots & \cdots & \cdots & \cdots & 0 & -1 & 2 \end{bmatrix}$$

where B, C, D, E, F are $(2k-2) \times (2k-2)$, $(2k-1) \times (2k-1)$, $(2k-1) \times (2k-1)$, $(2k-1) \times (2k-1)$, $(2k-1) \times (2k-1)$ respectively.

By using properties of determinants we get:

$$|B| = \begin{vmatrix} 2 & 0 & \cdots & \cdots & \cdots & \cdots & \cdots & \cdots & \cdots & \cdots & \cdots & \cdots & \cdots & 0 \\ 0 & 2 & 0 & \cdots & \cdots & \cdots & \cdots & \cdots & \cdots & \cdots & \cdots & \cdots & \cdots & \vdots \\ \vdots & \vdots & \frac{3}{2} & 0 & \cdots & \cdots & \cdots & \cdots & \cdots & \cdots & \cdots & \cdots & \cdots & \vdots \\ \vdots & \vdots & \vdots & \frac{4}{3} & 0 & \cdots & \cdots & \cdots & \cdots & \cdots & \cdots & \cdots & \cdots & \vdots \\ \vdots & \vdots & \vdots & 0 & \frac{3}{4} & 0 & \cdots & \cdots & \cdots & \cdots & \cdots & \cdots & \cdots & \vdots \\ \vdots & \vdots & \vdots & 0 & 0 & 2 & 0 & \cdots & \cdots & \cdots & \cdots & \cdots & \cdots & \vdots \\ \vdots & \vdots & \vdots & 0 & 0 & 0 & 2 & 0 & \cdots & \cdots & \cdots & \cdots & \cdots & \vdots \\ \vdots & \vdots & \vdots & 0 & 0 & 0 & 0 & \frac{3}{2} & 0 & \cdots & \cdots & \cdots & \cdots & \vdots \\ \vdots & \vdots & \vdots & \vdots & \vdots & \vdots & \vdots & 0 & \ddots & 0 & \cdots & \cdots & \cdots & \vdots \\ \vdots & \vdots & \vdots & \vdots & \vdots & \vdots & \vdots & \vdots & 0 & \ddots & \cdots & \cdots & \cdots & \vdots \\ \vdots & \vdots & \vdots & \vdots & \vdots & \vdots & \vdots & \vdots & \vdots & 0 & \frac{4}{3} & 0 & 0 & 0 & 0 \\ \vdots & \vdots & \vdots & \vdots & \vdots & \vdots & \vdots & \vdots & \vdots & \vdots & 0 & \frac{3}{4} & 0 & 0 & 0 \\ \vdots & \vdots & \vdots & \vdots & \vdots & \vdots & \vdots & \vdots & \vdots & \vdots & 0 & 0 & 2 & 0 & 0 \\ \vdots & \vdots & \vdots & \vdots & \vdots & \vdots & \vdots & \vdots & \vdots & \vdots & 0 & 0 & 0 & 2 & 0 \\ 0 & \cdots & \cdots & \cdots & \cdots & \cdots & \cdots & \cdots & \cdots & 0 & 0 & 0 & 0 & \frac{3}{2} \end{vmatrix} = 6^k$$

$$|D| = \begin{vmatrix} 2 & 0 & \cdots & \cdots & \cdots & \cdots & \cdots & \cdots & \cdots & \cdots & \cdots & \cdots & \cdots & 0 \\ 0 & 2 & \ddots & \cdots & \cdots & \cdots & \cdots & \cdots & \cdots & \cdots & \cdots & \cdots & \cdots & \vdots \\ \vdots & \ddots & \frac{3}{2} & \ddots & \cdots & \cdots & \cdots & \cdots & \cdots & \cdots & \cdots & \cdots & \cdots & \vdots \\ \vdots & \vdots & \ddots & \frac{4}{3} & \ddots & \cdots & \cdots & \cdots & \cdots & \cdots & \cdots & \cdots & \cdots & \vdots \\ \vdots & \vdots & \vdots & \ddots & \frac{3}{4} & \ddots & \cdots & \cdots & \cdots & \cdots & \cdots & \cdots & \cdots & \vdots \\ \vdots & \vdots & \vdots & \vdots & \ddots & 2 & \ddots & \cdots & \cdots & \cdots & \cdots & \cdots & \cdots & \vdots \\ \vdots & \vdots & \vdots & \vdots & \vdots & \ddots & 2 & \ddots & \cdots & \cdots & \cdots & \cdots & \cdots & \vdots \\ \vdots & \vdots & \vdots & \vdots & \vdots & \vdots & \ddots & \frac{3}{2} & \ddots & \cdots & \cdots & \cdots & \cdots & \vdots \\ \vdots & \vdots & \vdots & \vdots & \vdots & \vdots & \vdots & \ddots & \ddots & \ddots & \cdots & \cdots & \cdots & \vdots \\ \vdots & \vdots & \vdots & \vdots & \vdots & \vdots & \vdots & \vdots & \ddots & \frac{4}{3} & \ddots & \cdots & \cdots & \vdots \\ \vdots & \vdots & \vdots & \vdots & \vdots & \vdots & \vdots & \vdots & \vdots & \ddots & \frac{3}{4} & \ddots & \cdots & \vdots \\ \vdots & \vdots & \vdots & \vdots & \vdots & \vdots & \vdots & \vdots & \vdots & \vdots & \ddots & 2 & \ddots & \vdots \\ \vdots & \vdots & \vdots & \vdots & \vdots & \vdots & \vdots & \vdots & \vdots & \vdots & \vdots & \ddots & 2 & \ddots \\ \vdots & \vdots & \vdots & \vdots & \vdots & \vdots & \vdots & \vdots & \vdots & \vdots & \vdots & \vdots & \ddots & \frac{3}{2} & 0 \\ 0 & \cdots & \cdots & \cdots & \cdots & \cdots & \cdots & \cdots & \cdots & \cdots & \cdots & \cdots & 0 & \frac{-1}{3} \end{vmatrix} == 2*6^{k-1}$$

$$|E| = \begin{vmatrix} -1 & 0 & \cdots & \cdots & \cdots & \cdots & \cdots & \cdots & \cdots & \cdots & \cdots & \cdots & \cdots & 0 \\ 0 & -2 & \ddots & \cdots & \cdots & \cdots & \cdots & \cdots & \cdots & \cdots & \cdots & \cdots & \cdots & \vdots \\ \vdots & \ddots & -2 & \ddots & \cdots & \cdots & \cdots & \cdots & \cdots & \cdots & \cdots & \cdots & \cdots & \vdots \\ \vdots & \cdots & \ddots & \frac{3}{2} & \ddots & \cdots & \cdots & \cdots & \cdots & \cdots & \cdots & \cdots & \cdots & \vdots \\ \vdots & \cdots & \cdots & \ddots & \frac{4}{3} & \ddots & \cdots & \cdots & \cdots & \cdots & \cdots & \cdots & \cdots & \vdots \\ \vdots & \cdots & \cdots & \cdots & \ddots & \frac{3}{4} & \ddots & \cdots & \cdots & \cdots & \cdots & \cdots & \cdots & \vdots \\ \vdots & \cdots & \cdots & \cdots & \cdots & \ddots & -2 & \ddots & \cdots & \cdots & \cdots & \cdots & \cdots & \vdots \\ \vdots & \cdots & \cdots & \cdots & \cdots & \cdots & \ddots & \ddots & \ddots & \cdots & \cdots & \cdots & \cdots & \vdots \\ \vdots & \cdots & \cdots & \cdots & \cdots & \cdots & \cdots & \ddots & -2 & \ddots & \cdots & \cdots & \cdots & \vdots \\ \vdots & \cdots & \cdots & \cdots & \cdots & \cdots & \cdots & \cdots & \ddots & \frac{3}{2} & \ddots & \cdots & \cdots & \vdots \\ \vdots & \cdots & \cdots & \cdots & \cdots & \cdots & \cdots & \cdots & \cdots & \ddots & \frac{4}{3} & \ddots & \cdots & \vdots \\ \vdots & \cdots & \cdots & \cdots & \cdots & \cdots & \cdots & \cdots & \cdots & \cdots & \ddots & \frac{3}{4} & \ddots & \vdots \\ \vdots & \cdots & \cdots & \cdots & \cdots & \cdots & \cdots & \cdots & \cdots & \cdots & \cdots & \ddots & -2 & \ddots & \vdots \\ \vdots & \cdots & \cdots & \cdots & \cdots & \cdots & \cdots & \cdots & \cdots & \cdots & \cdots & \cdots & \ddots & -1 & 0 \\ 0 & \cdots & \cdots & \cdots & \cdots & \cdots & \cdots & \cdots & \cdots & \cdots & \cdots & \cdots & \cdots & 0 & -1 \end{vmatrix} == 2*6^{k-1}$$

$$|F| = \begin{vmatrix} 2 & 0 & \cdots & & & & & & & & & & & & 0 \\ 0 & 2 & \ddots & & & & & & & & & & & & \vdots \\ \vdots & \ddots & \frac{3}{2} & \ddots & & & & & & & & & & & \vdots \\ & & \ddots & \frac{4}{3} & \ddots & & & & & & & & & & \\ & & & \ddots & \frac{3}{4} & \ddots & & & & & & & & & \\ & & & & \ddots & 2 & \ddots & & & & & & & & \\ & & & & & \ddots & 2 & \ddots & & & & & & & \\ & & & & & & \ddots & \ddots & \ddots & & & & & & \\ & & & & & & & \ddots & \frac{3}{2} & \ddots & & & & & \\ & & & & & & & & \ddots & \frac{4}{3} & \ddots & & & & \\ & & & & & & & & & \ddots & \frac{3}{4} & \ddots & & & \\ & & & & & & & & & & \ddots & 2 & \ddots & & \\ & & & & & & & & & & & \ddots & 2 & \ddots & \vdots \\ \vdots & & & & & & & & & & & & \ddots & \frac{3}{2} & 0 \\ 0 & \cdots & & & & & & & & & & & & 0 & \frac{4}{3} \end{vmatrix} = \frac{4}{3} * 6^k$$

Therefore we get

$$\tau(S(F_3^k)) = \frac{1}{6^k} \begin{vmatrix} 5*6^{k-1} & 2*6^{k-1} \\ 2*6^{k-1} & \frac{4}{3}*6^k \end{vmatrix} = \frac{\frac{20}{3} \times 6^{2k-1} - 4 \times 6^{2k-2}}{6^k} = 6^k \qquad \blacksquare$$

## 4. Spanning tree entropy

The entropy of spanning trees of a network or the asymptotic complexity is a quantitative measure of the number of spanning trees and it characterizes the network structure. We use this entropy to quantify the robustness of networks. The most robust network is the network that has the highest entropy. We can calculate its spanning tree entropy which is a finite number and a very interesting quantity characterizing the network structure, defined as in [18] as:

$$Z(G) = \lim_{V(G) \to \infty} \frac{\ln \tau(G)}{|V(G)|} \qquad (1)$$

**Corollary** 4.1. The entropy of spanning trees of the friendship network $F_3^k$ is

$$Z(F_3^k) = 0.5493$$

**Proof**: From the Theorem 3. 2. and equation 1, and $|V(F_3^k)| = n = 2k + 1$ we obtain:

$$Z(F_3^k) = \lim_{n \to \infty} \frac{\ln 3^{(\frac{n-1}{2})}}{n} = \lim_{n \to \infty} \frac{(n-1)\ln 3^{(\frac{1}{2})}}{n} = \ln(\sqrt{3}) = 0.5493.$$

**Corollary** 4.2. The entropy of spanning trees of the friendship network $S(F_3^k)$ is

$$Z(F_3^k) = 0.3584$$

**Proof**: From the Theorem 3. 3. and equation 1, and $|V(F_3^k)| = n = 5k + 1$ we obtain:

$$Z(S(F_3^k)) = \lim_{n \to \infty} \frac{\ln 6^{(\frac{n-1}{5})}}{n} = \lim_{n \to \infty} \frac{(n-1)\ln 6^{(\frac{1}{5})}}{n} = \ln(\sqrt[5]{6}) = 0.3584.$$

**4. An Open Problem:**

Now, we suggest an open problem "Any improvement for calculating of the determinant in linear algebra, we can investigate this improvement as a new method to determine the number of spanning tree for a given graph.

## 4. Conclusion

In this paper, we contributed a new algebraic method to derive simple formulas of the complexity of some new networks using linear algebra. We applied this method to derive the explicit formulas for the friendship network $F_3^k$ and the subdivision of friendship graph $S(F_3^k)$. Finally, we calculated their spanning trees entropy and compare it between them.